\documentclass[prb,twocolumn,showpacs]{revtex4}

\usepackage{graphicx,epsf}
\usepackage{bm}

\begin{document}

\title{The Memory Effect in Electron Glasses
}
\smallskip

\author{Eran Lebanon and Markus M\"uller}
\affiliation{Center for Materials Theory, Serin Physics Laboratory,
    Rutgers University,\\
136 Frelinghuysen Road, Piscataway, New Jersey 08854-8019, USA}

\begin{abstract}
We present a theory for the memory effect in electron glasses. In fast gate 
voltage sweeps it is manifested as a dip in the conductivity
around the equilibration gate voltage. We show that this feature,
also known as anomalous field effect, arises from the long-time
persistence of correlations in the electronic configuration. We argue
that the gate voltage at which the memory dip saturates is related to an 
instability caused by the injection of a critical number of excess carriers. 
This saturation threshold naturally increases with temperature. On the other 
hand, we argue that the gate voltage beyond which memory is erased, is 
temperature independent. Using standard percolation arguments, we calculate 
the anomalous field effect as a function of gate voltage, temperature, carrier 
density and disorder. Our results are consistent with experiments, and in 
particular, they reproduce the observed scaling of the width of the memory 
dip with various parameters.
\end{abstract}


\maketitle

\section{Introduction}
\begin{figure}
\centerline{
\includegraphics[width=60mm]{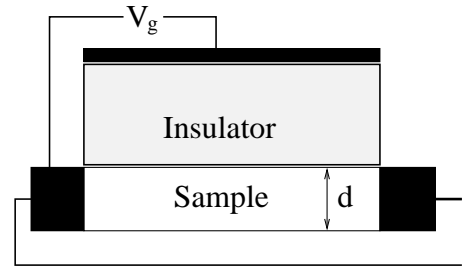}
} \vspace{-5pt} \caption{Typical experimental  setup: The sample
is a film (semiconductor or granular metal) of thickness $d$,
which is coupled capacitively to a gate electrode. Variation of the gate 
voltage $V_g$ slightly changes the number of carriers in the
film. The conductivity of the sample is probed through contact
electrodes that are directly attached to the sample.}
\label{fig:setup}
\end{figure}

In the insulating low temperature phase of dirty semiconductors or granular 
metallic films the unscreened Coulomb interactions between localized electrons 
lead to glassy behavior such as slow relaxation, history dependence of
observables, non-ergodicity, and memory effects. Even though such ``Coulomb 
glasses'' were theoretically predicted more than twenty years ago 
\cite{DaviesLee82,Gruenewald82,Davieslee84,Pollak84}, it was a major task 
to provide convincing experimental evidence for their existence. To our 
knowledge, very slow electronic relaxation was first reported in the context 
of capacitance measurements in doped GaAs by Monroe et al.~\cite{DonMonroe87}. 
At temperatures well below $1K$, they observed relaxation times that reached 
the scale of seconds. Even more striking non-equilibrium behavior was found by 
Ovadyahu's group in the conductivity of strongly disordered indium-oxide films 
\cite{benchorin93}, where the logarithmic relaxation can extend up to several 
hours (a typical experimental set-up is shown in Fig.~\ref{fig:setup}). Over the
last decade, careful studies of these systems have demonstrated that the 
electronic out-of-equilibrium behavior is indeed due to the strong frustration 
induced by the Coulomb interactions between localized electrons, and does not 
primarily reflect the glassy dynamics of extrinsic degrees of freedom
\cite{vaknin98}. All the key features usually associated with a glassy system 
have been observed in these systems: Aging\cite{vaknin00,orlyanchik04}, the dependence 
of sample properties on its history\cite{ovadyahu01,ovadyahu03} and memory effects 
\cite{vaknin02}. The latter appear as a dip in the film conductivity as the gate
voltage is swept through the point at which the system was equilibrated for a 
long time. The memory of these equilibration conditions usually persists for 
several hours after the gate voltage has been changed to a new value.

A very similar anomalous  field effect, accompanied by slow
relaxation, was observed in various granular metals such as 
Au~\cite{adkins84}, Al~\cite{Bielec01,Grenet03}, as well as Bi and 
Pb~\cite{MartinezArizala97,MartinezArizala98}. Furthermore, the aging
behavior and the temperature dependence of the memory dip reported
in granular Al~\cite{Grenet04} are very similar to those found in
 indium-oxide films~\cite{vaknin98b}. This suggests that these
glassy effects are rather universal, even though the details of
the hopping mechanism and the temperature dependence of the
resistivity are clearly different in the two systems.

The anomalous resistance fluctuations observed in ultrathin granular 
aluminium films~\cite{Bielec01} and silicon MOSFETs close to the metal 
insulator transition~\cite{Popovic02a,Popovic02b} were also interpreted 
as indications for glassy behavior. Unfortunately, in these systems, it is 
difficult to disentangle effects due to intrinsically glassy behavior of 
interacting electrons from the strong response of the 
percolating network of hopping electrons to extrinsic slow degrees of freedom. 

It has been conjectured that the glassy memory dip reflects a Coulomb 
gap~\cite{Pollak70,efrosshklovskii75} in the density of states which arises from the
unscreened Coulomb interactions. This conjecture is rather natural in the light of 
recent theoretical work which suggests that the emergence and the universality of 
the Coulomb gap are related to a glass transition at a finite $T_g$ in these 
systems~\cite{pastor99,MuellerIoffe04,pankov05}. This opens the appealing perspective 
of obtaining more detailed information on Coulomb correlations from a quantitative 
analysis of the memory experiments. 

So far, the precise connection between the memory effect and Coulomb correlations in the 
electron configuration has remained unclear. The aim of this paper is to provide a 
quantitative analysis of this effect, assuming a number of glassy properties of the 
electron system. We start with a review of the glassy features observed in experiments 
(Section~\ref{glassexp}), concentrating on the memory effect. In Section~\ref{Ingredients}, 
we briefly summarize the theoretical background on the Coulomb gap, hopping transport, 
percolation theory and glassy behavior which is needed for the quantitative theory of the 
memory effect in Section~\ref{theory}. The approximations underlying our theory and some open issues are 
discussed in Section~\ref{discussion}. We conclude 
with a brief summary of the main results. Several detailed discussions have been deferred 
to appendices in order not to interrupt the main line of the reasoning. 

\section{Glassy behavior in experiments}
\label{glassexp}
\subsection{Slow relaxation and aging}
In this section, we discuss some  of the experiments exhibiting
glassy behavior. Monroe et al.~\cite{DonMonroe87} were the first to observe slow electronic 
relaxation in the capacitance of
p-type doped, partially compensated GaAs. 
More recently, very slow logarithmic  relaxation was observed in
the conductivity of various granular
metals~\cite{MartinezArizala97,MartinezArizala98,Grenet03}, as
well as in indium-oxide films\cite{vaknin98}. Furthermore, after
equilibration under fixed experimental conditions and subsequent
moderate excitation during a time $t_w$ (e.g., by gate
voltage\cite{vaknin98} or electric field\cite{orlyanchik04}), the
typical relaxation time of such films is found to scale with $t_w$. This
phenomenon, known as simple aging, is observed in many glassy
systems such as polymers and supercooled liquids~\cite{Struick78}, as 
well as spin glasses~\cite{NordbladSvedlindh98}. 

\subsection{The memory effect}
\begin{figure}
\centerline{
\includegraphics[width=80mm]{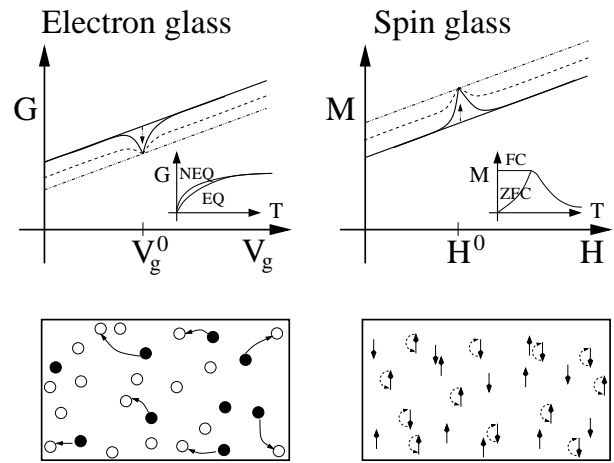} } \vspace{-5pt}
\caption{Left: In an electron glass, the conductivity exhibits a symmetric 
dip (on top of the linear normal field effect) around the equilibration gate 
voltage $V_g^0$. Far from $V_g^0$ the conductivity is higher
since a fast change of gate voltage takes the system into a higher-lying
metastable state. Below a characteristic temperature $T_g$, there is a
clear difference between the conductivity in an ``equilibrated'' state
(EQ) and metastable states (NEQ), see the inset. The relaxation from a
metastable state to an ``equilibrated'' state requires collective electron hops 
(lower left pannel) or crossing of high thermal barriers. Both are so slow that 
they cannot entirely take place on experimental time scales. However, partial 
relaxation results in the decrease of the dip amplitude with the sweep rate. \\
Right: An analogous experiment in spin glasses measures the magnetization as a 
function of magnetic field $H$ after a slow quench in the equilibration field $H^0$.
These traces exhibit a peak around $H^0$, reflecting that the field cooled
magnetization is larger than the magnetization that one obtains after a
quench in $H\neq H^0$ and a subsequent switch to $H^0$. The inset
illustrates this point with the well-known difference between field
cooled (FC) and zero-field cooled (ZFC) magnetization as a function of
temperature. The relaxation from the ZFC state to the FC state requires
the collective rearrangement of many spins (lower right pannel).}
\label{fig:EGvsSG}
\end{figure}

One of the most striking manifestations  of the electron glass is
the anomalous field effect: After the equilibration of the sample
at some gate voltage $V_g^0$, subsequent traces of
conductivity as a function of gate voltage keep a
long-lasting memory of these equilibrium conditions in the form of
a symmetric dip around $V_g^0$. This dip is superposed on the linear normal 
field effect due to the increase of carriers, which is usually subtracted 
and will not be considered further here. In Fig.~\ref{fig:EGvsSG}, we 
illustrate the memory effect - and for the reader more familiar with spin glasses, 
it is juxtaposed with an analogous experiment that could be done in spin systems.


The fact that the conductivity increases no matter whether carriers are added 
or depleted, can be understood on a qualitative level by the observation that 
any perturbation taking the system out of equilibrium must lead to an increase 
of the conductivity~\cite{benchorin93}. On a more quantitative level we will 
have to explain the following experimental observations extracted from extensive
studies on indium-oxide films~\cite{ovadyahu01}: 

\textit{(i)} The width $\Gamma$ of the memory dip (measured as a function of density of induced carriers) is 
remarkably universal.  In particular, it is independent of the sweep rate or the
application of a magnetic field. Even more surprisingly, it
remains unchanged under thermal annealing, a process which reduces
the disorder significantly and thus increases the conductivity by
several orders of magnitude. 

\textit{(ii)} $\Gamma$ increases with
carrier density. 

\textit{(iii)} $\Gamma$ increases roughly linearly with temperature, and keeps a certain memory 
of temperature: After a sudden quench from the equilibration temperature 
$T$ to a lower temperature $T'<T$, the width $\Gamma$ relaxes only slowly 
to the value corresponding to $T'$, keeping a memory of the larger width
characteristic of $T$\cite{vaknin02}. 

Note that \textit{(ii)} is a strong indication for the relevance of
electron-electron interactions\cite{vaknin98}. Similar results, in particular the 
decrease of $\Gamma$ with temperature and memory of higher
temperatures, were recently found in granular aluminium films,
too~\cite{Grenet03,Grenet04}.

We will show in Section~\ref{theory} 
that these key features 
can be understood semi-quantitatively with relatively simple arguments on the glassy free energy landscape and
the stability of its local valleys.

\section{Theoretical ingredients}
\label{Ingredients}

\subsection{Model}
\label{localizedmodel}
In Anderson insulators, the unscreened Coulomb interactions between the
localized carriers are the crucial ingredient which lead to strong
electron-electron correlations, the formation of the Coulomb gap and glassy behavior. 
An approximate description of such systems is given by the classical lattice Hamiltonian~\cite{efrosshklovskii75,efros76}
\begin{equation}
\label{Hamiltonian}
H=\sum_{i}n_i \epsilon_i +\frac{1}{2}\sum_{i,j}\frac{e^2}{\kappa}
\frac{n_in_j}{r_{ij}},
\end{equation}
where $n_i=0,1$ are the occupation numbers of randomly positioned lattice sites $i$, and we fix the 
average occupation to $1/2$. $\kappa$ denotes 
the host dielectric constant. The disorder energies $\epsilon_i$ are considered as 
independently and identically distributed random variables with a characteristic width $W$. This 
corresponds to a ``bare'' density of states $\nu_0 \sim 1/r_c^3W$ where $r_c=n_c^{-1/3}$ is the 
mean interparticle spacing in a system with carrier density $n_c$. 

Two standard models for semiconductors with localized electrons are usually considered in the 
literature~\cite{ESeedisorder,ESbook}: the``classical impurity band'' (CIB) and amorphous 
semiconductors (AS). They predominantly differ in the bare density of states $\nu_0$, see 
App.~\ref{semiconductors_models}: In the CIB the disorder $W$ is due to randomly 
distributed charged impurities, so that the disorder is of the order of the nearest neighbor 
interactions between carriers, and $\gamma\equiv W/(e^2/\kappa r_c) \sim 1$. In the AS, the disorder is due to strong local inhomogeneities, 
and can be much larger than the nearest neighbor interaction, $\gamma \gg 1 $. In indium-oxides, 
one can tune from the CIB-regime (low $n_c$) to the AS-regime (high $n_c$) by controlling the carrier density $n_c$,
 the crossover occurring around $n_X\approx \left(\frac{e^2/\kappa}{\hbar^2/2m} \right)^3$.

Most interesting glassy effects have been  observed in systems with a high density of carriers: 
in various granular metals and, most importantly, in indium-oxide. The latter is a highly disordered 
semiconductor, that can be prepared to have exceptionally high carrier densities in a large range 
$n_c\approx 10^{19}-10^{22}cm^{-3}$, while still being insulating. It is not obvious that these
high density systems can still be described by the lattice model~(\ref{Hamiltonian}), since 
typically the number of carriers per localization volume is larger than one. However, provided the 
parameter $z\equiv \nu_0\xi^{D-1} e^2/ \kappa$ is small ($\xi$ being the localization length), we can 
consistently restrict ourselves to a subsystem of well-separated electrons sitting on sites with small 
$\epsilon_i$, which maps the problem onto the effective model~(\ref{Hamiltonian}), as discussed in  App.~\ref{reduction}. 
We expect that the lattice model remains a reasonable approximation up to $z\approx 1$. Typical values for 
$z$ in indium-oxides can be estimated to be of the order of $z\approx 0.2-0.5$. 

\subsection{Coulomb gap and hopping transport}
\label{Coulombgapandhopping}
Since the pioneering works by Pollak\cite{Pollak70}, Efros and
Shklovskii\cite{efrosshklovskii75} in the early seventies, it has
been known that the unscreened Coulomb interaction in Anderson
insulators lead to important correlations in the configuration of
electrons, and in particular to the Coulomb gap in the density of states. 
An upper bound for the single particle density of states in $D$ dimensions is 
obtained from a self-consistent stability argument\cite{efrosshklovskii75,efros76}  
\begin{eqnarray}
\label{ESDOS}
\rho(E)\equiv \frac{1}{V}\sum_i\delta(E-E_i)\approx \left\{\begin{array}{ll}
\alpha_D\left(\frac{\kappa}{e^2}\right)^D E^{D-1},\ & E<E_C\\
\nu_0,\ &  E>E_C
\end{array}
\right.
\end{eqnarray}
where
\begin{equation}
E_C=(\left(e^2/\kappa\right)^D\nu_0/\alpha_D )^{ \frac{1}{D-1} }
\end{equation}
is the typical scale below which Coulomb correlations dominate
over the disorder, and $\alpha_D$ is a numerical constant. In (\ref{ESDOS}), 
$E_i\equiv dH/dn_i$ is the energy cost to change the occupation on the site $i$. 
A very similar distribution of energy costs (but with a substantially larger 
$\alpha_D$~\cite{Davieslee84,MuellerIoffe04}) holds for the quasiparticle 
excitations relevant for variable range hopping, sometimes referred to as 
electronic polarons~\cite{polaron,ESeedisorder}. At higher temperature, the 
Coulomb gap fills in gradually, and is essentially smeared out for $T>E_C$, even 
though a small depression due to Coulomb correlations should persist. 

The presence of a Coulomb gap at low temperatures leads to a crossover of the 
variable range hopping conductivity form Motts's law
\begin{equation}
\label{Mott} R(T) =  R_0 \exp\left\{ \left( \frac{T^{(D)}_M}{T}
\right)^{\frac{1}{D+1}}\right\},
\end{equation}
with $T_M^{(D)}\sim 1/\nu_0 \xi^D$ to the Efros-Shklovskii law
\begin{equation}
\label{ES} R(T) = R_0 \exp\left\{ \left(\frac{T^{(D) }_{ ES}}{ T}
\right)^{1/2} \right\},
\end{equation}
with $T_{ES}^{(D)}\sim e^2/\kappa\xi$. The prefactor $R_0$ is a slowly varying function of 
temperature. 

One can obtain a quantitative description of variable range hopping from the standard application of percolation theory to a
network of Miller-Abrahams resistors formed by pairs of
sites~\cite{Ambegaokar71,ShklovskiiEfros71,Pollak72,EfrosNguyen79}, as reviewed in
App.~\ref{percolationtheory}. 
In particular, considering an equilibrium quasiparticle density of states of the form
(\ref{ESDOS}) one finds a crossover function
\begin{equation}
\label{crossover}
\log[R(T)/R_0]=z^{-1/(D-1)}{\cal R}(T/T_X),
\end{equation}
where
\begin{equation}
\label{TX}
T_X=(T_{ES}^{D+1}/T_M^2)^{1/(D-1)}\sim z^{1/(D-1)} E_C
\end{equation}
is the temperature where the conductivity crosses over from Mott's regime to the 
Efros-Shklovskii regime (see Refs.~\cite{Aharony92,meir96} for similar approaches). 
In Section~\ref{theory}, we will be concerned with the modification of the conductivity as 
the density of states is driven out of equilibrium by the application of a gate voltage.

For a long time, experimental evidence for the Coulomb gap in doped 
semiconductors was only indirect in the form of the Efros-Shklovskii 
hopping law, and it is difficult to extract detailed information on Coulomb 
correlations from the temperature dependence of the resistivity $R(T)$ alone. 
However, \textit{differential} measurements 
represent direct fingerprints of correlations since they are only sensitive
to \textit{changes} in the electron configuration as an external
parameter is varied. 

In the last ten years, several tunneling experiments on weakly insulating samples provided direct evidence for a pseudogap in the density of states around the Fermi level~\cite{massey95,lee99,teizer00,sandow01,butko00,Butko00b,Butko01,lee05}. However, such experiments are restricted to the regime relatively 
close to the metal-insulator transition. From this point of view, the analysis of the anomalous field effect represents a convenient method to probe Coulomb correlations also deeper in the insulating regime. 

\subsection{Glassiness}
A key ingredient to the understanding of the memory effect is the glassiness of 
the electrons at low temperatures.
Such a behavior can be expected 
from the theoretical consideration that Coulomb systems are very similar to frustrated 
antiferromagnets, for which recent experiments have demonstrated the existence of a 
thermodynamic phase transition and spin-glass-like out of equilibrium behavior in almost pure
samples~\cite{Ladieu04}.

From numerical simulations, it is well established that at sufficiently low temperatures there is a 
multitude of metastable states, which are not ergodically connected within timescales accessible 
in a simulation, since they are separated by large activation or tunneling 
barriers ~\cite{perez-garrido99,menashe00,menashe01,Tsigankov03,Grempel04,Kolton05}. Extrapolating 
to experimentally relevant timescales, one expects that a dynamical glass transition takes place 
in real systems as well. 

More theoretical insight can be gained from mean-field theory~\cite{MuellerIoffe04,pankov05}. Let 
us first discuss the case of strong disorder ($\gamma>1$), which corresponds to most experiments 
in indium-oxide ($n_c\gtrsim n_X$).  In 3D systems, a locator approximation to the high temperature expansion
predicts a glass transition at finite temperature~\cite{MuellerIoffe04} 
\begin{eqnarray}
\label{Tg3DlargeW}
T_g^{(3D)}&=&\frac{1}{6(2/\pi)^{1/4}} \alpha_3^{ 1/2 } E_C,
\end{eqnarray}
valid for large disorder, $\gamma\gg 1$. 
Applying the same approach to 2D systems in the strong disorder
limit, one obtains the prediction
\begin{eqnarray}
\label{Tg2DlargeW}
T_g^{(2D)}&=&\frac{\sqrt{8\pi}}{\log\left(\frac{\sqrt{\pi/2}}{z}\right)}
\alpha_2 E^{(2D)}_C.
\end{eqnarray}

It is possible that in systems in low dimensions the sharp mean field thermodynamic transition 
is rounded due to activation over finite but high barriers. In this event, $T_g$ in 
Eqs.~(\ref{Tg3DlargeW},\ref{Tg2DlargeW}) is expected to mark 
the crossover to strongly activated dynamics. 

Apart from predicting $T_g$, mean field theory tells us that in 2D the ratio $T_g/E_C$ (\ref{Tg2DlargeW}) 
can be numerically large, in particular, if we remember that the relevant value of $\alpha_2$ is the one 
associated with the quasiparticle density of states. 
One may therefore expect glassy behavior in strongly disordered films in a substantial range of
temperatures $T/E_C>1$ where the Coulomb gap is hardly developed yet.
It is indeed not unusual that the glass transition occurs at a temperature where the density of states does not yet exhibit any of its prominent low temperature features. This is for example the case in the Sherrington-Kirkpatrick model of long range spin glasses where a linear pseudogap starts to open only well below $T_g$. The same is predicted by the mean field solution for strongly disordered Coulomb glasses.

In the case of moderate disorder, $\gamma\approx 1$ (the CIB model) 
mean field theory predicts a rather low glass transition temperature~\cite{pankov05} in 3D,
\begin{eqnarray}
\label{Tg3Dimp}
T_g^{(3D)}&\approx& 0.03  \frac{e^2 n_c^{1/3}}{\kappa},
\end{eqnarray}
consistent with the small values found in simulations on irregular lattices without on-site
disorder~\cite{GrannanYu93, OverlinYu04}.
We expect a similar situation in weakly disordered 2D systems.

In indium-oxide films, a rapid drop of
the relaxation time is observed as the carrier density is decreased below $n_{cr}
\approx 10^{19}cm^{-3}$, while keeping the temperature constant~\cite{vaknin98}. It is possible that this is a manifestation of the glass transition. 
Indeed, such samples are in the classical impurity band regime ($n_{cr}<n_X$), and Eq.~(\ref{Tg3Dimp}) yields a value of $T_g$
close to the measurement temperature of $T_m=4K$. 
Films with even lower density, $n_c<n_{cr}$, are still in their ergodic high
temperature phase at $T=T_m$, and the relaxation times are unmeasurably
fast. 

\section{Theory of the memory dip}
\label{theory}
\subsection{The density of states as a function of $T$ and $V_g$}
Our theoretical approach is based on the assumption that the metastable states visited after an excitation by gate voltage reflect the way in which this state was reached. In particular, we argue that at a new gate voltage $V_g$ the (quasiparticle) density of states, and thus the hopping conductivity, will be distinct from the equilibrated state at the same $V_g$, even if all spontaneous single-particle relaxations had time to take place. The full relaxation to equilibrium will involve multi-particle relaxations and/or processes with high activation energies. Analytical and numerical arguments in support of this scenario have been discussed in Ref.~\cite{MullerLebanon05}.

Here we are focusing on truly 2D systems, that is, films of thickness $d\lesssim d_{cr}\equiv  (\nu_0 e^2/\kappa)^{-1/2}$ (see App.~\ref{dimensionality} for a discussion of the crossover to 3D systems).
As discussed in Section~\ref{Coulombgapandhopping},  the
 density of states in such films exhibits a linear Coulomb gap at low
temperatures. At energies larger than the Coulomb correlation scale 
$E_C= (e^2/\kappa)^2 \nu_0^{(2D)}/\alpha_2$, the density of
states approaches the constant bare density of states
$\nu_0^{(2D)}=\nu_0d$. We describe this crossover (for
$T=0$) by the interpolating function
\begin{eqnarray}
\rho_0(E) = \nu_0^{(2D)} \ {\rm tanh} | E/E_C |,
\end{eqnarray}
whose precise form is, however, not essential for the following analysis.
\begin{figure}
\centerline{
\includegraphics[width=80mm]{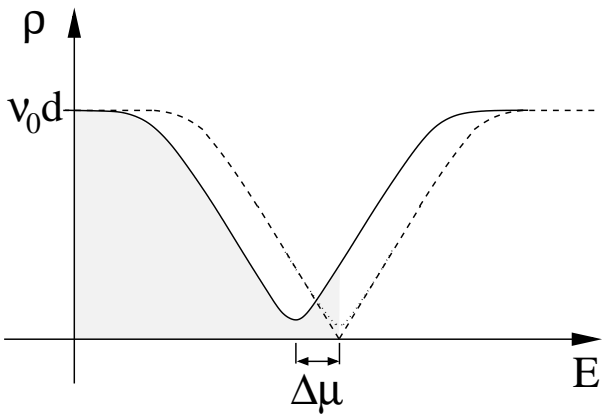}
}\vspace{-5pt} \caption{A sketch of the density of states in a 2D electron
glass.  The dashed curve corresponds to zero temperature. The dotted
curve shows the result of thermal smearing, cf.~Eq.(\ref{rhoofT}).
The solid  curve represents the density of states immediately
after applying a gate voltage, cf.~Eq.(\ref{rho_of_V}). $\Delta \mu$ is 
the shift of the chemical potential due to the charging of the sample. }
\label{fig:smearDOS}
\end{figure}

\begin{figure}
\centerline{
\includegraphics[width=80mm]{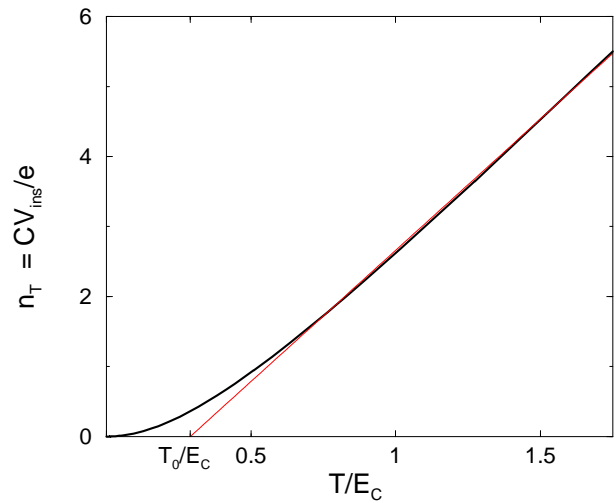}
}\vspace{-5pt} \caption{(Color online) The density of thermally excited
carriers, $n_T$, as a  function of temperature, or, equivalently,
the gate voltage difference $CV_{\rm ins}/e$ at which the anomalous
field effect saturates. At high temperature, $V_{\rm ins}$ is
linear in $T$ [$\propto (T-T_0)$], and scales as $T^2$ at low temperatures. $n_T$ and
$C V_{\rm ins}/e$ are plotted in units of $zE_C\nu_0 d$.}
\label{fig:width_vs_T}
\end{figure}

At finite temperature some electrons are excited out of their
local equilibrium position, which induces fluctuations in the (single 
electron) site energies, $E_i = \epsilon_i+ \sum_{j\ne i}e^2n_j/  \kappa 
r_{ij}$:
\begin{equation}
E_i =E_i^{(0)}+\delta \phi_i
\end{equation}
where
\begin{equation}
E_i^{(0)}=\epsilon_i+ \sum_{j\ne i} e^2n_j^{(0)}/   \kappa r_{ij}
\end{equation}
is the energy cost to change the occupation of site $i$ in the locally stable configuration characterized by $\{n_i^{(0)}\}$, and
\begin{equation}
\delta \phi_i=\sum_{j\ne i} e^2 \delta n_j/  \kappa r_{ij}
\end{equation}
are potential fluctuations due to thermally activated  changes in
the occupation $\delta n_j$. In a first approximation we assume
the $\delta \phi_i$'s to be independent Gaussian distributed  
variables with variance $\langle \delta \phi ^2 \rangle = \alpha_T
n_T (e^2/\kappa)^2$,
\begin{equation}
P_T(\delta\phi) = \frac{\kappa}{e^2}\frac{1}{\sqrt{\pi\alpha_T  n_T}} \exp
\left\{ -\frac{\kappa^2}{e^4}\frac{\delta\phi^2}{ \alpha_T n_T} \right\},
\end{equation}
where $\alpha_T=O(1)$ is a numerical factor and $n_T$ is the 
density of thermally excited electrons
\begin{equation}
n_T = \int_{-\infty}^0 dE \ \rho_0(E)  \left( 1-f(E)   \right) +
\int_0^{\infty} dE \ \rho_0(E) f(E),
\end{equation}
$f$ being the Fermi distribution. Note that at  high
temperatures $n_T$ is linear in $T$, while at low temperatures it
approaches zero as $n_T \sim T^2$ (see Fig.~\ref{fig:width_vs_T}).

As a consequence of these fluctuations, the
density of states
is smeared. In the approximation of independent shifts
$\delta \phi_i$, it is described by the convolution
\begin{equation}
\label{smearT}
\rho(E,T)  = \int d(\delta \phi) P_T(\delta  \phi) \rho_0( E -
\delta\phi). \label{rhoofT}
\end{equation}

Upon application of a gate voltage $V_g$, new carriers are introduced into the system. It is 
reasonable to assume them to rapidly occupy the empty sites with the lowest energies $E_i$ 
available. Sometimes, the occupation of a site may cause a small number of neighboring particles 
to hop slightly away, to reduce the energy of the system. Considering the introduction of 
the new particle and such local rearrangements as one composite process, one may say that the 
new carriers actually occupy ``quasiparticle'' states (very similar to the electronic polarons relevant for conductivity~\cite{polaron}). 
However, even though the thus reached state may be stable to single particle relaxations, it will 
in general be an excited state under the new gate voltage. Only if the system is given a long time 
to equilibrate, it will relax to the new ground state, which involves multiparticle transitions or 
the crossing of high activation energies. 

In order to describe the quasiparticle density of states analytically, we assume that the latter 
type of relaxation processes has not had time to occur. Furthermore, we assume that apart from the 
local response of nearby particles, the introduction of new carriers does not 
trigger major rearrangements of the electron configuration. This will be justified further below. A 
more thorough investigation of these assumptions can be found in \cite{MullerLebanon05}.

Under these assumptions, the effects of the gate 
voltage are twofold: ({\em i}) the new carriers successively fill the empty
(quasiparticle) levels close to the Fermi energy, shifting the chemical potential to $\mu+\Delta\mu$, 
while the minimum in the density of states
remains at the old value of $\mu$ (cf.~Fig.~\ref{fig:smearDOS}); ({\em ii}) the extra particles
further smear the density of states, similarly to the thermal
effect described above. They induce further energy shifts $\delta
\phi_i$, which we take to be randomly distributed according
to
\begin{equation}
P_{V_g}(\delta \phi) = \frac{ \kappa}{e^{2}}\frac{1}{\sqrt{ \pi \alpha_V
CV_g/e}} \exp \left\{ -\frac{\kappa^2}{e^4}\frac{(\delta \phi - \Delta
\mu)^2}{\alpha_V CV_g/e} \right\},
\end{equation}
where $\alpha_V=O(1)$ and $C$ is  the capacitance per unit area.
We have also accounted for the global shift in chemical
potential $\Delta \mu$, which is related to the gate voltage by
\begin{eqnarray}
CV_g = e\int_0^{ \Delta \mu(V_g)} dE \rho(E) .
\end{eqnarray}
Notice that in the presence  of a Coulomb gap, the
dependence of $\Delta \mu$ on $V_g$ is non-linear. The 
density of states after a sudden gate voltage change is finally
obtained as
\begin{equation}
\label{rho_of_V} \rho(E,T,V_g) = \int d(\delta\phi)
P_{V_g}(\delta \phi) \rho (E-\delta\phi , T, V_g=0).
\end{equation}
The density of states at different stages of smearing is shown in
Fig~\ref{fig:smearDOS}. 
Below, we will use the density of states 
(\ref{rho_of_V}) to calculate quantitatively the out of equilibrium 
conductivity

Note that in assumption $(i)$ it is implied that new carriers will occupy sites
across the whole film. This is only justified if the film thickness is of the order of the screening length of the sample. In the absence of a Coulomb gap the latter can be estimated as $l_{sc}\sim (\nu_0 e^2/\kappa)^{-1/2}$ which is of the same order as the thickness $d_{cr}$ which governs the crossover to a 3D system. (However, in the presence of an Efros-Shklovskii Coulomb gap, the screening length is probably substantially larger on intermediate timescales, as suggested by the capacitance experiments of Ref.~\cite{DonMonroe87}.)
We believe that in films thicker than $l_{sc}$, the anomalous field effect is mostly due to the filling of states within a screening length from the surface. The quantitative theory below does not strictly apply to this case. However, the instability argument given in the following subsection should still hold, provided the film thickness $d$ is replaced by $d_{cr}\sim l_{sc}$.

\subsection{Instability criterion and breakdown of memory}
The description~(\ref{rho_of_V}) of an adiabatic response  to the change of
gate voltage $V_g$, without any relaxation of the electron
configuration, is applicable only for small enough values of
$V_g$. As the gate voltage is increased, more and more new
particles are introduced into the  sample and reshuffle the site
energies, until at a certain scale ($V_g=V_{\rm ins}$) the local 
minimum in which the system resides becomes unstable. 
For higher gate voltages the system will relax 
to a new local minimum whose density of states is no longer 
described adequately by the adiabatic  smearing and shifting of 
Eq.~(\ref{rho_of_V}) alone.  As long as the Coulomb gap is not 
strongly developed, it is reasonable to assume that the new local 
minimum represents a rather generic metastable state relatively high up in the energy spectrum of all possible states. 
We then expect that the conductivity will not significantly 
change upon further increase of the gate voltage, since the 
system remains in the high energy spectrum of
states. However, such a new metastable 
state will 
still have a large configurational similarity (or ``overlap'' in 
spin glass language) with the original ground state: Most of the
sites that were occupied in the original local state remain
occupied in the new local minimum. The memory of the original
configuration is thus preserved. In particular, when the gate
voltage is swept back to its original value, the low equilibrium
conductivity will be recovered. As the gate voltage is increased
beyond the instability scale $V_{\rm ins}$, the overlap of the new 
local minimum with the original state continuously decreases and the memory 
of the original state is gradually lost. This will be manifested by
the disappearance of the memory dip once the gate voltage
is swept beyond a scale $V_{\rm mem}$. 

\begin{figure}[t]
\centerline{
\includegraphics[width=80mm]{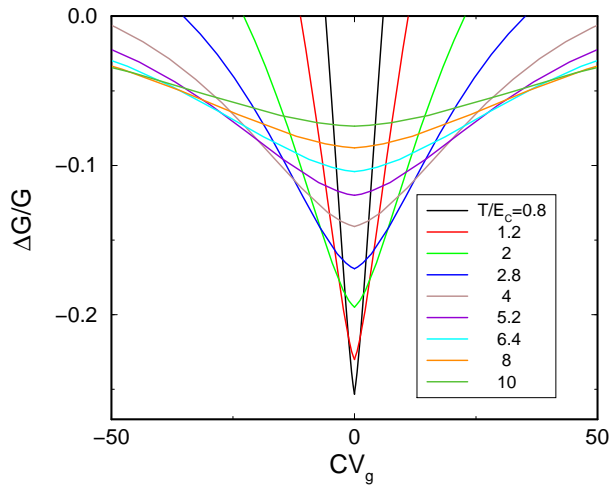}
}\vspace{-5pt} \caption{(Color online) Memory dip as a function  of gate voltage
for $z=0.4$ and various temperatures of the order of $E_C$  and higher.
We plot the relative change in conductivity with respect
to its asymptotic value at large gate voltages, $\Delta G/G=[
G(V_g)-G(V_{\rm ins}) ]/G(V_{ \rm ins})$. The cusp width is
proportional to temperature, and its amplitude decreases as
temperature is increased. $C V_g$ is plotted in units of
$ezE_C\nu_0d$} \label{fig:cusps_high_T}
\end{figure}

Let us now analyze in more detail what determines  the instability
scale $V_{\rm ins}$. It is reasonable to expect that as long as
the density of carriers introduced by the gate is smaller than the
density of thermally excited electrons $n_T$, the gate voltage
effect is perturbative, which justifies our adiabatic treatment of
the density of states. This reasoning implies $CV_{\rm ins}
\gtrsim en_T$.

On the other hand, at gate voltages $CV_g> en_T$  the shift in
chemical potential is of the order of the temperature $\Delta \mu
\sim T$, and accordingly, the new carriers are introduced on sites
that were essentially empty in the original state. The
local environment of those sites will generally not be favorable to the addition of a
new particle. Rather, the newly introduced electron will trigger fast
relaxation processes and destabilize the original state. In
other words, the configuration generated by occupying more and more levels
 will soon become a generic high energy state
for $CV_g> en_T$. In summary, we expect an instability and thus a saturation of the out-of-equilibrium conductivity at
\begin{eqnarray}
\label{Vins}
V_{\rm ins} \approx \frac{e n_T}{C} \approx \frac{e \nu_0 d}{C} \times
\left\{
\begin{array}{lcr}
 \pi^2T^2/6E_C, &\ & T \ll E_C \\ \\
2{\rm ln}2\ T-\Omega/2\nu_0d, &\  & T>E_C
\end{array}
\right.
\end{eqnarray}
where $\Omega=\int dE (\nu_0d-\rho(E))\sim \nu_0 d  E_C$ is the
total deficit of density of quasiparticle states due to the presence of a Coulomb gap. In
Fig.~\ref{fig:width_vs_T} we plot $V_{\rm ins}$ as a function of temperature.

As explained above, we expect the memory dip to saturate around the 
instability scale, its total width being roughly $\Gamma =2V_{\rm ins}$. A
linear high temperature behavior $\Gamma \sim T-T_0$ was indeed observed
experimentally~\cite{private_com_Cusp}. Equation (\ref{Vins}) further predicts
the interesting relation $T_0=\int dE (1-\rho(E)/\nu_0d)/4\ln 2$,
which allows to determine experimentally the width of the Coulomb
gap.

The memory of the equilibrium  state is
essentially erased once the gate voltage has exceeded a certain scale
$V_{\rm mem}>V_{\rm ins}$. We expect this crossover to occur when the
random energy shifts $\delta \phi_i$ are comparable to the Coulomb correlation scale, $\langle \delta\phi^2\rangle^{1/2}\approx E_C$. More explicitly, we
obtain the estimate
\begin{equation}
\label{Vmem}
V_{\rm mem} \approx \zeta_V \frac{e\nu_0 d}{C} E_C,
\end{equation}
which is temperature independent contrary to $V_{\rm  ins}$. The
numerical factor $\zeta_V=O(1)$ is presumably numerically large (at least of
the order of $T_g/E_C$ which may be appreciable in strongly disordered 2D systems, cf.~(\ref{Tg2DlargeW})). 

In Ref.~\cite{ovadyahu03} the authors reported that
the ratio $V_{\rm mem}/V_{\rm ins}$ is not a universal number (at
fixed temperature), but increases with carrier density. The above
arguments indeed suggest that  at high temperatures, $T\gtrsim E_C$,
\begin{equation}
\frac{V_{\rm mem}}{V_{\rm ins}}\propto\frac{E_C}{T-T_0}.
\end{equation}
At fixed temperature, this ratio increases with carrier 
concentration, as $E_C$ does.

\begin{figure}[t]
\centerline{
\includegraphics[width=80mm]{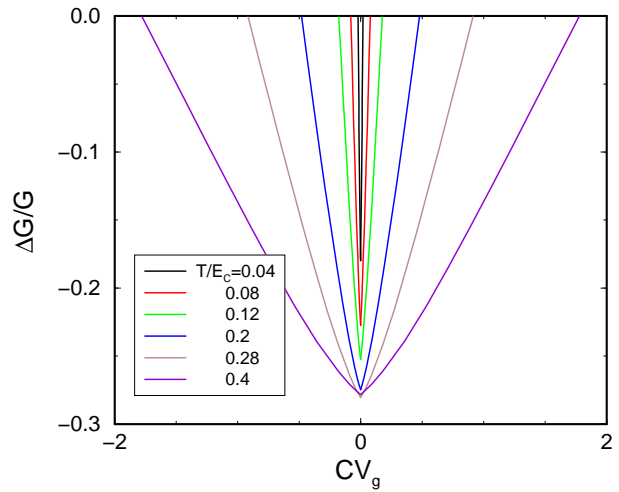}
}\vspace{-5pt} \caption{(Color online) $\Delta G/G=[ G(V_g)-G(V_{\rm ins})
]/G(V_{ \rm ins})$ plotted against gate voltage for low
temperatures, $T<E_C$, and $z=0.4$. In this regime, the cusp width
increases quadratically with temperature. The adiabatic
percolation treatment predicts the amplitude of the cusp to
increase with temperature. This is probably an artifact, see the
discussion in Section~\ref{discussion}. $CV_g$ is plotted in
units of $ezE_C\nu_0d$.} \label{fig:cusps_low_T}
\end{figure}

\subsection{Non-equilibrium conductivity}
\label{noneq} 
For gate
voltages $V\lesssim V_{\rm ins}$ we can calculate the non-equilibrium
conductivity from the modified density of states,
Eq.~(\ref{rho_of_V}), and the percolation criterion of
App.~\ref{percolationtheory}. Assuming that the conductivity
saturates to $G_\infty$ beyond the scale $V_{\rm ins}$, we may
estimate $G_\infty \approx G(V_g=V_{\rm ins})$ from which we
obtain the amplitude of the memory dip as $\Delta G\equiv
G_\infty-G(0)\approx G(V_{\rm ins})-G(0)$.

The smearing of the density of states due to new
carriers has a minor effect on the non-equilibrium conductivity
since it mostly affects energy scales on the order of the
temperature whereas the energy range probed by variable range
hopping is much larger. However, the shift of the chemical
potential, $\Delta \mu$ is the crucial out-of-equilibrium feature which leads to the increase of the conductivity. 

\begin{figure}
\centerline{
\includegraphics[width=80mm]{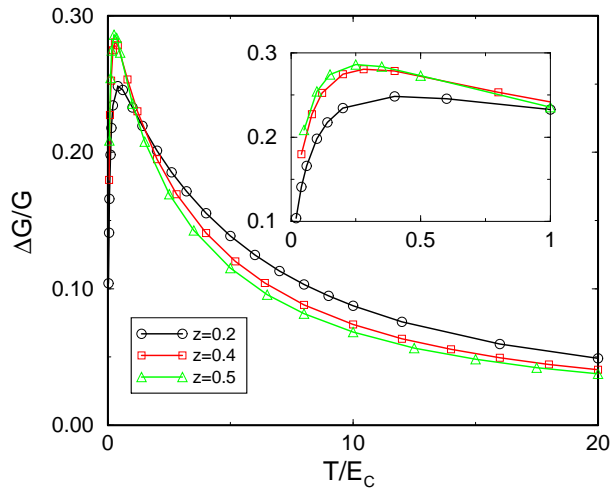}
}\vspace{-5pt} \caption{(Color online) The relative amplitude of the conductivity
dip $[G(V_{\rm ins})-G(0)]/G(V_{\rm ins})$ as a function of
temperature for various values of $z$. Inset: Low temperature
behavior of the relative amplitude for the same values of $z$.}
\label{fig:reldip}
\end{figure}

The results we obtained from the percolation treatment confirm the
general assertion~\cite{vaknin00} that the conductivity always increases with
$|V_g|$. 
In Fig.~\ref{fig:cusps_high_T} we plot the relative change of the conductivity $\Delta
G(V_g)/G_\infty$ for high temperatures $T_g>T\gtrsim E_C$.
This is the temperature regime in which most experiments on indium-oxide 
films are performed. The cusp width decreases linearly with temperature
while its amplitude increases, an asymptotic analysis yielding
\begin{equation}
\label{dGGhighT}
\frac{\Delta G}{G} \sim \frac{E_C}{T}, 
\end{equation}
see~Fig.~\ref{fig:reldip}.

At low temperature, $T<E_C$,
the cusp width increases quadratically with $T$. In
contrast to the high temperature regime, the adiabatic percolation
treatment combined with the instability criterion predict $\Delta G/G_\infty$ to increase with
temperature as
\begin{equation}
\label{dGGlowT}
\frac{\Delta G}{G} \sim \sqrt{\frac{T}{T_{ES}}}.
\end{equation}
The full functions $\Delta G(V_g)/G_\infty$ are shown in Fig.~\ref{fig:cusps_low_T}. 

\begin{figure}[t]
\centerline{
\includegraphics[width=80mm]{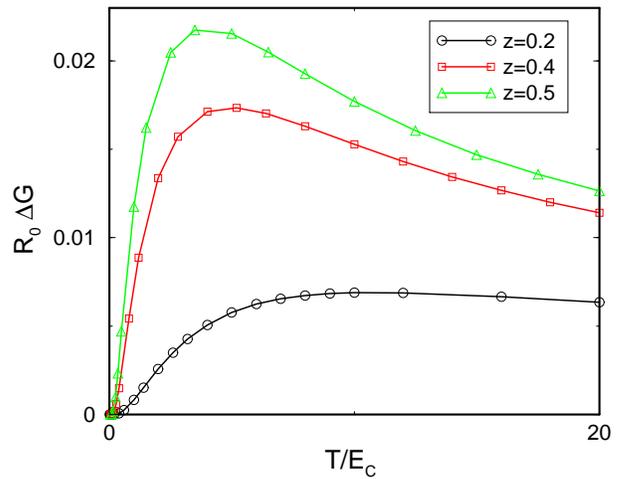}
}\vspace{-5pt} \caption{(Color online) Amplitude of the  conductivity dip
$G(V_{\rm ins}) -G(0)$, as a function of temperature for $z=0.2$,
$0.4$ and $0.5$. Less resistive films (larger $z$) exhibit
 a clear maximum at relatively high temperatures.
}
\label{fig:dip_vs_T}
\end{figure}

The \textit{absolute} amplitude of the dip,  $\Delta G$, is
found to be a non-monotonic function of temperature, as shown in Fig.~\ref{fig:dip_vs_T}. The non-monotonicity is more
pronounced in less resistive films (see the curves for the localization parameters $z=0.4,\ 0.5$) 
where a clear peak appears in $\Delta G$ at $T_{\rm max}$, while for more
resistive films ($z=0.2$) this feature is hard to discern. Very
similar non-monotonic behavior of $\Delta G$ was observed
experimentally~\cite{vaknin98b}. 

\section{Discussion}
\label{discussion}

\subsection{Low temperature behavior}
\label{dis_shortcomings} 
The non-monotonic behavior of the {\em relative}
amplitude predicted by Eqs.~(\ref{dGGhighT},\ref{dGGlowT}) was not observed in experiments so far. Even though this might be
due to the fact that most experiments are performed at high
temperatures, we believe the
prediction $\Delta G/G\rightarrow0$ for $T\rightarrow 0$ to be an artifact for 
the following reasons: ({\em i}) One can trace back the origin of the 
prediction $\Delta G/G\sim T^{1/2}$ to the conservation of the total deficit of density of states under the application of a gate voltage. In our 
approximation, this is a simple consequence of only shifting and convoluting 
the density of states. However, it is likely that for $V\sim V_{\rm ins}$ the neglected fast 
relaxation processes destroy this exact conservation, which would lead to the saturation 
of $\Delta G/G$ at low temperatures. ({\em ii}) The assumption that the 
asymptotic conductivity $G_{\infty}$ is well estimated by the
non-equilibrium conductivity at $V_{\rm ins}$ is probably
incorrect at very low temperatures. In order to illustrate this point, let us consider a large gate
voltage $V_L$ with $ V_{\rm ins}\ll V_L\leq V_{\rm mem}$, such that the 
shift of the chemical potential is of the order of $E_C$. This will
take the system into a high energy state where fast relaxation
processes lead to the formation of a new Coulomb gap. The sites within the 
new gap will be mostly different from the sites in the old one, and it is 
likely that the total deficit of density of states in this new Coulomb gap is initially smaller than the one in 
the equilibrium state. More precisely, one may expect that the linear
slope $\alpha_{ES}^{2D}$ of the density of states in the new 
configuration is slightly larger than that in the ground state, and the corresponding
Efros-Shklovskii temperature $T_{ES}$ is smaller accordingly (see 
App.~\ref{percolationtheory}). This would suggest that $\Delta G/G$ 
 scales like $T^{-1/2}$ and thus \textit{increases} with decreasing
temperature. In this temperature regime, one may expect an
additional increase of conductivity with increasing gate voltage,
even beyond the instability point $V_{\rm ins}$. The description of the
conductivity in that regime would require to take into
account partial relaxation processes which goes beyond the present
approach. However, it remains open whether the $\sqrt{T}$ behavior Eq.~(\ref{dGGlowT}) applies in an
intermediate range of temperatures.

\subsection{Comparison to experiments}
\label{dis_rep}
\subsubsection{The memory dip}
As described in Section~\ref{noneq}, the adiabatic
percolation approach reproduces well the temperature dependence of
several key features of the memory dip. In particular, we showed
that one
may infer the width of the quasiparticle Coulomb gap from a careful study of the 
temperature dependence of the width. Moreover, from Eq.~(\ref{Vins}) we 
see that the width of the memory dip should be proportional to the bare 
density of states $\nu_0 $, which increases with carrier concentration as
discussed in Section~\ref{localizedmodel}: In the impurity band
regime, $n_c<n_X$, one expects $\nu_0 \sim \ n_c^{2/3}$, while for large 
carrier concentration the density of states crosses over to a free 
electron-like behavior $\nu_0 \sim n_c^{1/3}$. This scenario agrees
rather well with the observed bending of the dip width as a
function of carrier density in indium-oxide films~\cite{vaknin98}.

At high temperatures, the percolation approach predicts a decrease
of the cusp amplitude like $1/T$, which is weaker than  what is
usually observed in experiments~\cite{vaknin98b,Grenet03}. This
difference may again originate from our neglect of spatial
correlations in $\delta \phi_i$ discussed above. Even more likely
is the scenario that our assumption of homogeneously glassy
samples breaks down at higher temperatures. Indeed, if only a few
rare regions with stronger disorder remain glassy their effect on
the out of equilibrium conductivity might be strongly reduced due
to shortcuts by non-glassy regions.

Let us briefly discuss the effect of varying the disorder
strength or applying a magnetic field. Experimentally, both changes seem 
not to affect the width of the memory dip. Furthermore, even strong magnetic 
fields change the amplitude $\Delta G/G$ only slightly. These experimental 
observations are quite surprising since both the variation of disorder and 
magnetic field affect the conductivity $G$ itself appreciably. 

The first observation can be naturally explained within our picture if
we assume that changing the disorder (by annealing) or 
applying a magnetic field mostly affects the
localization length without altering the bare
density of states $\nu_0$. Since the instability criterion originates
from a static consideration within the classical model Hamiltonian 
(\ref{Hamiltonian}), it is insensitive to the localization length. 
The width of the memory dip remains thus constant under variations of 
disorder or magnetic field.

The near constancy of $\Delta G/G$ with magnetic field is more
subtle. In terms of the percolation approach the variation of
the localization length simply changes the parameter $z$, while
leaving $E_C$ fixed. From Fig.~\ref{fig:reldip}, it can be seen
that the effect of $z$ on $\Delta G/G$ is indeed relatively small,
while the corresponding change of $\Delta G$ (cf.~Fig.~\ref{fig:dip_vs_T}) is much more important.

\subsubsection{Memory of temperature}
An interesting effect of temperature memory was reported both for
the indium-oxide films~\cite{vaknin02}, and for films of granular
aluminum~\cite{Grenet04}. After equilibration at $V_g=0$ and
temperature $T_0$, the system is quenched to $T_1<T_0$, and
the conductivity is probed as a function of gate voltage before 
the sample equilibrates. In this protocol, the anomalous field 
effect maintains the characteristic width of the initial temperature 
$T_0$ for a rather long time before narrowing down. 

This effect can be understood in terms of the instability criterion proposed 
above: The energy minima or valleys in which the electron glass 
typically settles at temperature $T_0$ will be stable upon injection 
of additional carriers up to the critical density $n_{T_0}$. After a 
temperature quench the system will in general remain in this valley, 
and the stability threshold of the higher temperature will be preserved 
temporarily. Finally, slow relaxation processes allow 
the system to dig itself deeper down into a subvalley of the energy 
landscape, since the thermal fluctuations are now smaller. These lower 
lying states will have a reduced stability threshold, as will be reflected 
by a smaller dip width $\Gamma$.

\subsection{Observability of glassy effects in doped semiconductors}
So far, slow electronic relaxation was observed only in very few cases of moderately doped semiconductors~\cite{DonMonroe87}. The natural question thus arises as to why
glassy effects, such as in indium-oxide, are not more frequently
encountered. The reason is most likely that many standard
semiconductors have relatively low carrier concentrations with well
localized electrons. Such systems are described by the classical
impurity  band model whose glass transition in 3D is suppressed by
a small numerical factor, see Eq.~(\ref{Tg3Dimp}). Estimating the
Efros-Shklovskii conductivity at that temperature scale, one
finds
\begin{equation}
 ln(R/R_0)\approx \left(C_{ES}^{(3D)}\frac{e^2}{\kappa \xi
T_g}\right)^{1/2}\approx
\left(\frac{C_{ES}^{(3D)}}{0.03}\frac{n_c^{-1/3}}{\xi}\right)^{1/2},
\end{equation}
which is very large even when the localization length approaches
the inter-impurity distance.
This makes the detection of glassy effects in the hopping conductivity 
nearly impossible, due to the intrinsically large noise in such 
systems. However, glassiness should still be observable in static 
quantities, such as the capacitance measurements of 
Ref.~\cite{DonMonroe87}.

In amorphous semiconductors with relatively high carrier concentration 
glassy effects as described in this paper should generally be observable. 
The same is true for doped semiconductors sufficiently close to the metal 
insulator transition. 

Once the localized wavefunctions start to overlap significantly 
($z>1$) one may expect the nature of the glass phase to change and finally 
disappear completely. Many recent experiments probing the Coulomb gap are 
actually carried out in this regime~\cite{Butko01,helgren04,lee05}. They 
reveal very interesting quantum critical behavior associated with the metal 
insulator transition, but have not thoroughly investigated the glassy aspects 
of the samples so far. First attempts towards a theoretical description of 
glassiness in the regime close to the transition were undertaken in 
Refs.~\cite{dalidovich02,dobrosavljevic03}. At this point it remains 
an interesting open question, both theoretically and experimentally, whether 
the onset of metastability and glassiness coincides with the transition to 
the insulator, and, if so, what role the glassy freezing plays in the physics 
of the metal-insulator transition. 

\section{Conclusion}
\label{conclusion} 
We have analyzed the memory effect in electron glasses. The 
non-equilibrium conductivity was calculated within a percolation 
approach, assuming the local metastability of the glass state. This allowed us to describe the anomalous field effect quantitatively, reproducing many of the 
experimental characteristics observed in indium-oxides and granular
aluminum. We have provided a simple physical picture for the voltage scales 
at which the memory dip saturates and erasure of memory occurs, respectively.
We argue that the saturation scale increases with temperature, its dependence on temperature reflecting the characteristics of the Coulomb gap. We have predicted the ratio of the two voltage scales as a function of temperature and carrier density, which can be tested in experiments. 

\section{Acknowledgments}
We acknowledge discussions with P. Chandra, M. Feigel'man, M. Gershenson, 
T. Grenet, L. Ioffe, Z. Ovadyahu and B.I. Shklovskii. We thank L. Ioffe and 
Z. Ovadyahu for the continuous encouragement and interest in our work, as 
well as for the frequent exchange of ideas. E.L. was supported by DOE grant
DE-FE02-00ER45790. M.M. was supported by NSF grant DMR 0210575.

\appendix

\section{The bare density of states of semiconductors}
\label{semiconductors_models}
In this appendix we derive approximate expressions for the bare density of states in semiconductors and granular metals.

In the literature, two standard models for
semiconductors with localized electrons have been considered, see
Refs.~\cite{ESeedisorder,ESbook} for a review. The
``classical impurity band'' model refers to lightly doped,
partially compensated semiconductors where all carriers are
localized within a Bohr radius around majority impurities. Due to the Coulomb interactions with randomly
distributed charged impurities, the on-site energies $\epsilon_i$
of these localized states are scattered over a range of the order
of the nearest neighbor interactions,
$e^2/\kappa r_c$, where $r_c\equiv n_c^{-1/3}$ is the  average distance between
carriers, $n_c$ is the carrier density (uncompensated dopant
concentration) and $\kappa$ is the host dielectric constant. Accordingly, the bare
density of localized states (which neglects Coulomb interactions
between the localized carriers) is of the order of
\begin{equation}
\label{DOSimp}
\nu_0=\frac{n_c}{e^2/\kappa r_c}=\frac{\kappa}{e^2}n_c^{2/3}, \quad
\textrm{(impurity band)}.
\end{equation}

\begin{figure}
\centerline{
\includegraphics[width=60mm]{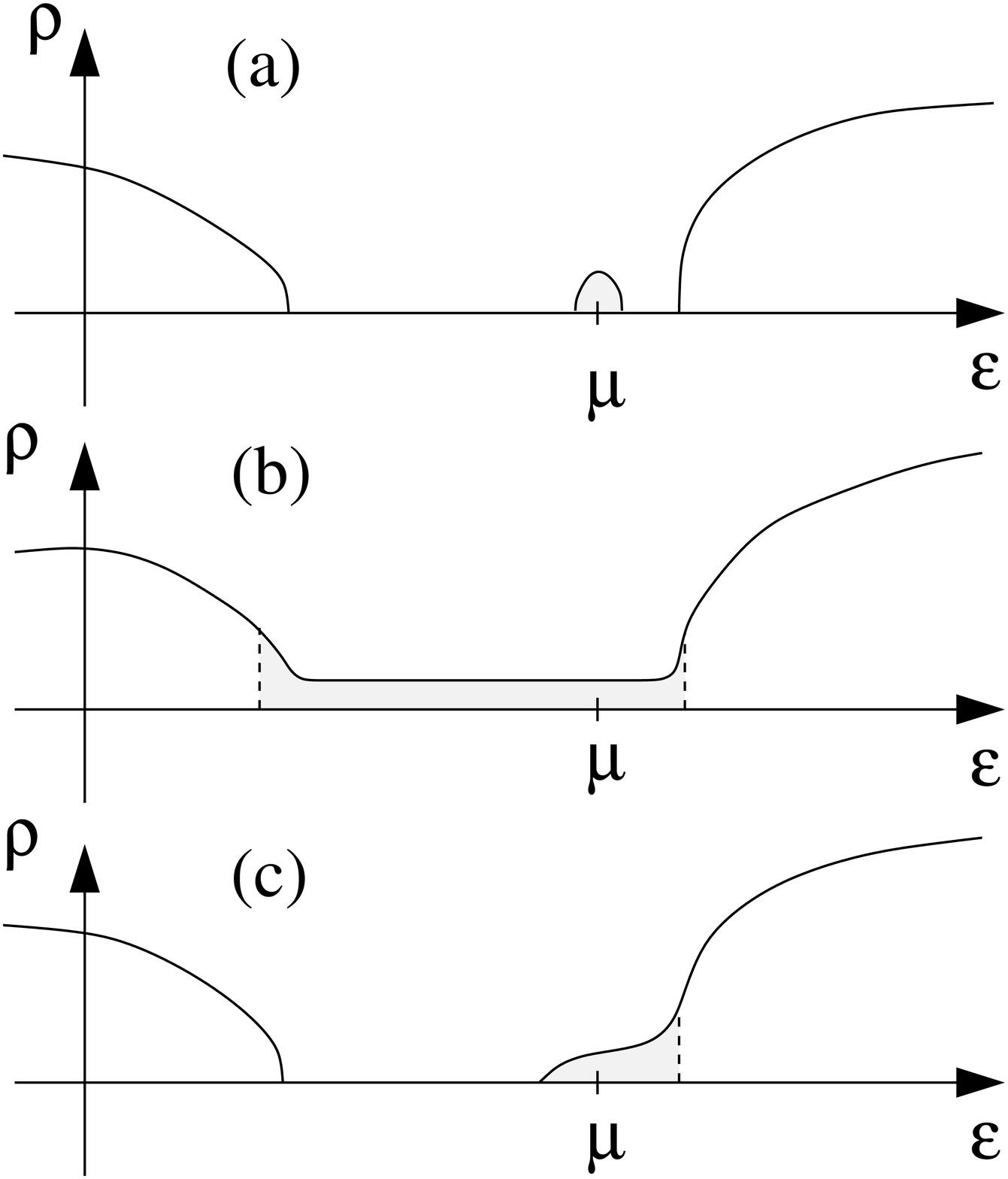}
}\vspace{-5pt} \caption{Schematic view  of the bare density of
states (neglecting electron-electron interactions) in different
classes of semiconductors: a) classical impurity band (lightly
doped semiconductors) b) strongly disordered amorphous
semiconductor (e.g., amorphous germanium), c) doped, disordered
semiconductor with localized band tails at the bottom of the
conduction band (e.g., indium-oxide).} \label{fig:DOSinSC}
\end{figure}

The second frequently considered model  describes amorphous
semiconductors in which the disorder of the on-site energies
$\epsilon_i$ is due to strong local inhomogeneities. In this case,
their scatter is usually much larger than that introduced by
Coulomb interactions with impurities. As a consequence, the
bare density of states is lower than (\ref{DOSimp}), as
schematically illustrated in Fig.~\ref{fig:DOSinSC}.

The set of localized states  does not need to fill the
whole region between the valence and the conduction band. In
the case of indium-oxide (both amorphous~\cite{Ovadyahu93} and
crystalline~\cite{CohenOvadyahu94}), it has been established that
the localized states form a tail joining the conduction band at the
mobility edge. Furthermore, it was found that at sufficiently high
carrier densities, the density of states in the range of localized
states is in surprisingly good agreement with free-electron
estimates,
\begin{equation}
\label{freeelDOS}
\nu_0 \approx \frac{n_c}{\hbar^2 n_c^{2/3}/2m}, \quad
\textrm{(high density, $n_c>n_X$).}
\end{equation}
This reflects the fact that the kinetic  energy $E_{kin}=\hbar^2 k_F^2/2m 
\sim \hbar^2 n_c^{2/3}/2m$ of the localized wavefunctions
dominates over the effects of inhomogeneities in the electrostatic
potential. Note, however, that a crossover to the regime of dominant
Coulomb interactions (Eq.~(\ref{DOSimp})) is to be expected around $n_c\approx n_X$ where $E_{kin}\approx e^2n_c^{1/3}/\kappa$, i.e.,
\begin{equation}
n_X\approx \left(\frac{e^2/\kappa}{\hbar^2/2m} \right)^3.
\end{equation}

In granular metals, the role of the localized  sites in the model
(\ref{Hamiltonian}) is taken by the grains. As a consequence of
impurities and the disorder in the size and the arrangement of the
grains, the cost to introduce one more particle on a grain is a
random quantity of the order of the typical charging energy $E_C$
or the level spacing $\delta$ in the grain, whichever is larger.
Usually the charging energy will dominate, unless the grains are
very small or the effective dielectric constant of the metallic
film is very large. The random on-site energies $\epsilon_i$
entering the Hamiltonian (\ref{Hamiltonian}) are therefore
scattered with a typical width $W=\max[E_C, \delta]$, and the
effective bare density of states can be estimated as
\begin{equation}
\label{grainDOS} \nu_0\approx \frac{n_c}{\max[E_C,\delta]},
\quad\textrm{(granular metals)}
\end{equation}
which is typically  a few times smaller than the (2D) density of
states in a bulk metal.

\section{Reduction of high density systems to the standard model}
\label{reduction}
Here we examine under which condition a high density system can be described by the classical Hamiltonian (\ref{Hamiltonian}). 
The aim is to consider only a strip of localized states of width
$\Delta E$ around the chemical potential, and to work with an
effective model of occupied and empty levels within this
strip. In this
approximation, the carriers localized in states of lower energy
are considered inert in the sense that they do not hop to other
sites. Notice however, that such ``core'' electrons may still have
fairly extended wavefunctions, and therefore contribute to the
polarizability of the medium, renormalizing the host dielectric
constant.

The mapping of such an energy strip to a  model of point-like localized
states (\ref{Hamiltonian}) is consistent provided
that (\textit{i}) the states within the strip do not overlap spatially,
\begin{equation}
\label{strip1}
\Delta E \nu_0 \xi^D<1,
\end{equation}
and that (\textit{ii}) the typical  variations $\delta \phi$ of the
electrostatic energy due to rearrangements of particles within the
strip do not exceed the width $\Delta E$ of the strip,
\begin{equation}
\label{strip2} \delta \phi \sim \left(\nu_0\Delta
E\right)^{1/D}\frac{e^2}{\kappa}< \Delta E.
\end{equation}
The conditions (\ref{strip1}) and (\ref{strip2})  can be
satisfied simultaneously if
\begin{equation}
\label{z}
z = \frac{e^2}{\kappa}\nu_0 \xi^{D-1}<1,
\end{equation}
or, in other words, if the level spacing within a  localization
volume is larger than the Coulomb interaction strength on the
scale of the localization length. 

\section{The crossover from 2D to 3D}
\label{dimensionality} 
In this appendix we discuss the crossover thickness $d_{cr}$  below 
which a sample should be considered two-dimensional. In particular, we show that 
the crossover from a bulk sample to a film occurs around a thickness
\begin{equation}
\label{dcr}
d_{cr}\sim (\nu_0 e^2/\kappa)^{-1/2},
\end{equation}
both with respect to transport characteristics and the glass transition.
[Using typical values for indium-oxide films ($\nu_0\approx
10^{32}erg^{-1}cm^{-3}$ and $\kappa \approx 30$~\cite{private_com_Cusp}) one finds
$d_{cr}\sim 100 \AA$ which is of the order of the typical film
thickness ($d=50-200 \AA$) in most glassy experiments.]

\begin{figure}
\centerline{
\includegraphics[width=60mm]{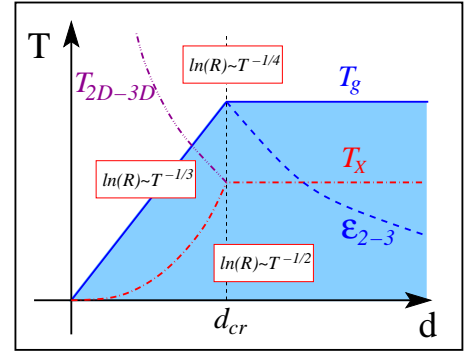}
}\vspace{-5pt} \caption{(Color online) Phase diagram of
electron glasses as a function of temperature and film thickness.
The dash-dotted lines labeled $T_X$ red (light grey) and $T_{2D-3D}$ purple (dark grey) indicate the crossover between different hopping
regimes. The solid lines separate the ergodic high temperature
phase from the glass phase where memory effects and aging are
observable. The dashed line indicates the temperature and energy
scale below which the density of states assumes a linear shape
characteristic for two dimensions.} \label{fig:Dimcrossover}
\end{figure}

In a thin film, Mott's variable range  hopping law (\ref{Mott})
crosses over from the 3-D form ($\log(R)\sim T^{-1/4}$) to the 2-D
form ($\log(R)\sim T^{-1/3}$) when the hopping length becomes of
the order of the film thickness, which yields the crossover
temperature
\begin{equation}
T_{2D-3D}\sim T_M^{(3D)} (\xi/d)^4\sim  T_X^{(3D)}/(\nu_0 d^2
e^2/\kappa)^4.
\end{equation}
A subsequent crossover to the Efros-Shklovskii  law takes place at
$T_X^{(2D)}= T_X^{(3D)}(\nu_0 d^2 e^2/\kappa)$, where $T_X^{(3D)}$
denotes the Mott to Efros-Shklovskii crossover temperature for a
bulk sample. The intermediate regime with a 2D-Mott's law is
observable only if $d<d_{cr}$.

The crossover from a 3D to a 2D glass transition  occurs when the
typical distance between thermally active sites at $T=T_g^{(3D)}$
becomes equal to the film thickness, i.e., when
\begin{equation}
R_{T_g}\sim e^2/\kappa T_g^{(3D)}\sim e^2/\kappa T_g^{(2D)}\sim d.
\end{equation}
One can easily check that these expressions all  become of the same
order when $\nu_0 d^2 e^2/\kappa= (d/d_{cr})^2\approx 1$. Notice
that for films with $d<d_{cr}$, the glass transition temperature
(\ref{Tg2DlargeW}) decreases roughly linearly with thickness since
$T_g^{(2D)}\sim \nu_0^{(2D)}\sim d \nu_0$. The phase diagram
 as a function of temperature and film
thickness is summarized in Fig.~\ref{fig:Dimcrossover}.

\section{Percolation theory of hopping conductivity}
\label{percolationtheory}

In this appendix we review the percolation theory of hopping conductivity.
We consider the network of Miller-Abrahams resistors  formed by
pairs of sites $i$ and $j$. In the vicinity of a given low
temperature metastable state of the electron glass, the effective
resistance of this link is approximately given by
\begin{eqnarray}
R_{ij}\approx R_0\exp\left(-2r_{ij}/\xi+\epsilon_{ij}/T\right)
\end{eqnarray}
where
\begin{eqnarray}
\epsilon_{ij}=\left\{ \begin{array}{ll}
|E_i-E_j|-e^2/\kappa r_{ij}, & \textrm{if  }E_i\cdot E_j<0\\
\max\left\{|E_i|,|E_j|\right\}, & \textrm{if  }E_i\cdot E_j>0
\end{array}\right.
\end{eqnarray}
and $E_i$ is the energy (with respect to the chemical  potential)
to remove or add a particle at the site $i$ in the particular
metastable state at hand. More precisely, the energies $E_i$ refer
to the excitation of quasiparticles (or polarons~\cite{polaron}) that carry the
hopping current.

In order to find the least resistive percolating path in the resistor
network we follow the procedure proposed by Efros et
al.~\cite{EfrosNguyen79}: We consider only resistors with
$R_{ij}<R_0\exp(\chi_c)$ to be active and associate to each of them a
disk or ball with diameter $r_{ij}$. We finally determine the
threshold value of $\chi_c$ for which the set of disks percolates. The value $R_0 \exp(\chi_c)$ 
is a good estimate of the resistivity to exponential
accuracy.

To solve this problem analytically, one needs to  know the
probability $F(\omega,r)$ per unit energy and volume to find a
pair of sites $(i,j)$ with $r_{ij}=r$ and $\epsilon_{ij}= \omega$.
Under the assumption that the site energies $E_i$ are
independently distributed according to a single-quasiparticle
density of states $\rho(E)$, we obtain
\begin{eqnarray}
\label{F_omega_r} F(\omega,r)=\frac{1}{2}\int \rho(E_1)\rho(E_2)
\delta(\epsilon_{12}-\omega)dE_1dE_2.
\end{eqnarray}
With the help of the pair distribution function the above percolation problem reduces to
that of a set of balls with different radii. Assuming that the
critical volume fraction of the balls, $\Theta_D$, is an
approximate invariant of temperature, only dependent on
dimensionality, we finally have to solve the equation
\begin{eqnarray}
\label{percolationcriterion} \Theta_D = \int d\omega d^Dr
V_D\left(\frac{r}{2}\right)^D
F(\omega,r)\theta(\chi_c-\frac{\omega}{T}-\frac{2r}{\xi}),
\end{eqnarray}
where $V_D$ is the volume of a $D$-dimensional  unit sphere. The
above invariance principle yields $\Theta_2\approx
1.26$\cite{Nguyen84} and $\Theta_3\approx
0.23$\cite{EfrosNguyen79}). Comparison with other percolation
criteria, in particular in the Mott
regime~\cite{SeagerPike74,PikeSeager74,ESbook}, indicate that in
$2D$ a slightly smaller value $\Theta_2\approx 1$ yields results
closer to the numerically found percolation threshold. In the main
part of the paper we therefore used the latter value.

In order to efficiently implement the percolation criterion for an  arbitrary density
of states, as obtained, e.g., after the sudden application of a gate
voltage, it is convenient to introduce the functions
\begin{eqnarray}
\label{Fph}
F_{ph}(E)=2 \int_0^E d\epsilon \rho(\epsilon)\rho(\epsilon-E)\\
F_{pp}(E)=\rho(E)\int_0^E d\epsilon \rho(\epsilon)\\
F_{hh}(E)=\rho(-E)\int_{-E}^0 d\epsilon \rho(\epsilon)
\end{eqnarray}
and $\Phi_{\alpha\beta}(E)=\int_0^Ed\epsilon F_{\alpha\beta}
(\epsilon) $ with $\alpha,\beta\in \{p,h\}$, in terms of which the
percolation criterion (\ref{percolationcriterion}) can be
rewritten as
\begin{eqnarray}
\label{percolation}
\Theta_D&=&\int_0^{\xi\chi_c/2}d^Dr V_D \left( \frac{r}{2} \right)^D  \cdot\\
&&\left[ \Phi_{pp}(T(\chi_c-2r/\xi)) +\Phi_{hh}(T(\chi_c-2r/\xi)) \right.
\nonumber\\
&&\left.+\Phi_{ph}(T(\chi_c-2r/\xi)+e^2/\kappa r)-\Phi_{ph}(e^2/\kappa r)
\right].\nonumber
\end{eqnarray}

For completeness we report the standard expressions that  one
obtains in the limiting case of a constant density of states
($\rho(\epsilon)\equiv \nu_0$) and high temperatures ($T\gg T_X$,
Mott regime), and in presence of an Efros-Shklovskii pseudogap,
$\rho(\epsilon)= \alpha_D (\kappa/e^2)^D \epsilon^{D-1}$,
($T<T_X$). In these cases the above criterion is readily evaluated
analytically and yields the threshold values
\begin{eqnarray}
\label{Mottasymptotics}
\chi_c&=&\left(\frac{T^{D}_M}{T}\right)^{1/(D+1)},\,\, \textrm{(Mott)}\\
\chi_c&=&\left(\frac{T^{D}_{ES}}{T}\right)^{1/2},\,\, 
\textrm{(Efros-Shklovskii)}
\label{ESasymptotics}
\end{eqnarray}
with
\begin{eqnarray}
\label{Tchar}
T_M^{(D)}&=&\frac{C_M^{(D)}}{\nu_0 \xi^D}\\
T_{ES}^{(D)}&=&C_{ES}^{(D)}\frac{e^2}{\kappa\xi}.
\end{eqnarray}
The numerical constants $C^{(D)}_{M}$ depend on the value  of
$\Theta_D$, while $C^{(D)}_{ES}$ increases with the value of
$\Theta_D/\alpha_D^2$. In the main text we used $\Theta_2=1$.

\bibliographystyle{prsty}

\begin{thebibliography}{10}
\bibitem{polaron} ``Polaron'' excitations consist in the change of occupation on one site, and a small number of short hops of neighboring particles that decrease the energy cost for the local charge excitation~\cite{ESeedisorder}.
\end{thebibliography}



\end{document}